\documentclass{MolSpaLab}
\usepackage{graphicx}

\begin{document}
\TitreGlobal{Molecules in Space \& Laboratory}
\title{Millimeter wave spectral line surveys and line mapping studies of NGC6334I and I(N)}
\author{\FirstName S. \LastName Thorwirth}
\address{Max-Planck-Institut f\"ur Radioastronomie, Auf dem H\"ugel 69, 53121 Bonn, Germany}
\author{\FirstName A.J. \LastName Walsh}
\address{Centre for Astronomy, James Cook University, Townsville, QLD 4810, Australia}%
%
\author{\FirstName F. \LastName Wyrowski\,$^1$}
\author{\FirstName P. \LastName Schilke\,$^1$}

\author{\FirstName H. \LastName Beuther}
\address{Max-Planck-Institut f\"ur Astronomie, K\"onigstuhl 17, 69117 Heidelberg, Germany}
\author{\FirstName T.R. \LastName Hunter}
\address{National Radio Astronomy Observatory, 520 Edgemont Road, Charlottesville, VA 22903, U.S.A.}
\author{\FirstName C. \LastName Comito\,$^1$}
\author{\FirstName S. \LastName Leurini}
\address{ESO, Karl Schwarzschild Str. 2, 85748 Garching, Germany}
\author{\FirstName A.R. \LastName Tieftrunk}
\address{Im Acker 21b, 56072 Koblenz, Germany}
\author{\FirstName M.G. \LastName Burton}
\address{School of Physics, University of New South Wales, Sydney, 2052 NSW, Australia}
\author{\FirstName K.M. \LastName Menten\,$^1$}

%
\runningtitle{Spectra line surveys and mapping of NGC6334 I and I(N)}
\setcounter{page}{1}

\maketitle
\begin{abstract}
NGC6334I and I(N) have been observed with the Swedish-ESO Submillimetre Telescope, SEST,
at wavelengths of 3, 2, and 1.3\,mm. Especially 
NGC6334 I shows rich emission from many different molecules, 
comparable in line density to prototypical hot cores such as 
Orion-KL and SgrB2(N).
In addition, a 4$^\prime\times 4^\prime$ region enfolding NGC\,6334 I and I(N) has been mapped at a wavelength of 3\,mm
(75 to 116\,GHz) with the Mopra telescope.
\end{abstract}
%
\section{Introduction}
Along a gas filament of 11 pc, the southern giant molecular cloud NGC6334 exhibits several luminous sites of massive star formation
with source I dominating the millimeter to the far infrared (Sandell 2000).
In close proximity to NGC6334I, about two arcminutes to the north, another bright source is found 
in the millimeter- and submillimeter continuum, denoted NGC6334I(N)(e.g., Gezari 1982).
This source is believed to be a comparably young object
and there is compelling evidence
that star formation is going on there (Megeath \& Tieftrunk 1999, Persi et al. 2005).
It is this twin core system, NGC6334I and I(N), that offers the rare opportunity to study the evolution of high mass stars from the same parental cloud and in a relatively small spatial region.

To investigate the physical and chemical properties of NGC6334 I and I(N) in more detail,
we have started concerted campaigns making use of both single dish telescopes
and interferometers. ATCA investigations of NH$_3$
emission up to the (6,6) inversion transition reveals the presence of warm gas in both cores
(Beuther et al. 2007) and SMA continuum observations at 1.3\,mm (Hunter et al. 2006) resolve each core into a sample of sub-cores of several tens of solar masses each, nicely demonstrating the formation of star clusters.

Single dish molecular line observations show both cores to be chemically very rich (e.g., Thorwirth et al. 2003, Schilke et al. 2006). SEST surveys obtained at 2\,mm are shown in Fig. 1.
In addition, Mopra line maps of a 4$^\prime\times 4^\prime$ region enfolding NGC\,6334 I and I(N) have been obtained at 3\,mm covering nineteen different molecules. Sample maps of $^{13}$CO($1-0$), HCN($1-0$), and N$_2$H$^+$($1-0$) are shown in Fig. 2.
 A detailed analysis of the spectral line surveys and line mapping studies will be given elsewhere in due course.

\begin{figure}[ht]
\begin{center}
\includegraphics[width=11.5cm]{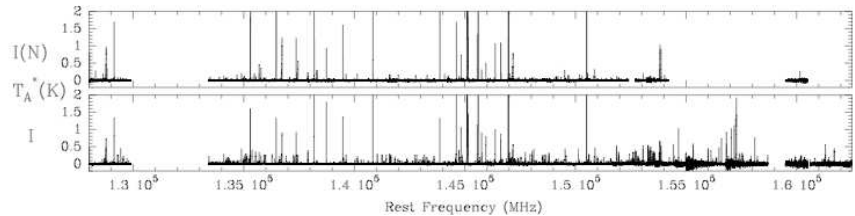}
\caption{SEST 2mm line surveys towards NGC6334 I (bottom) and I(N) (top). } \label{fig}
\end{center}
\end{figure}

\vspace*{-.5cm}
\begin{figure}[!h]
  \centering
  \begin{minipage}[t]{3cm}
\begin{center}
\hspace*{-.75cm}
    \includegraphics[scale=0.225]{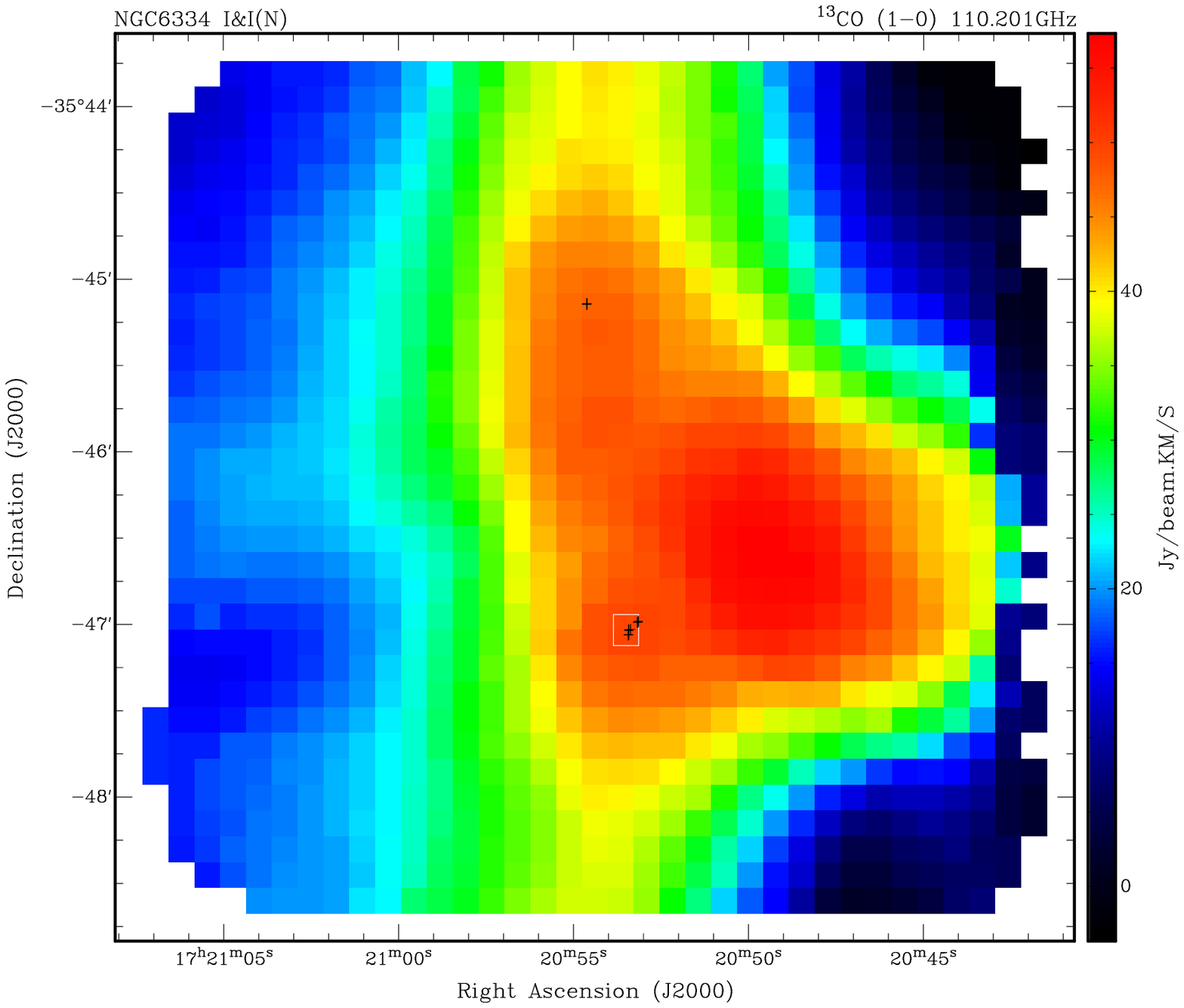} 
\end{center}
  \end{minipage}
  \hspace*{1.1cm}
  \begin{minipage}[t]{3cm}
\begin{center}  
\hspace*{-.75cm}
      \includegraphics[scale=0.225]{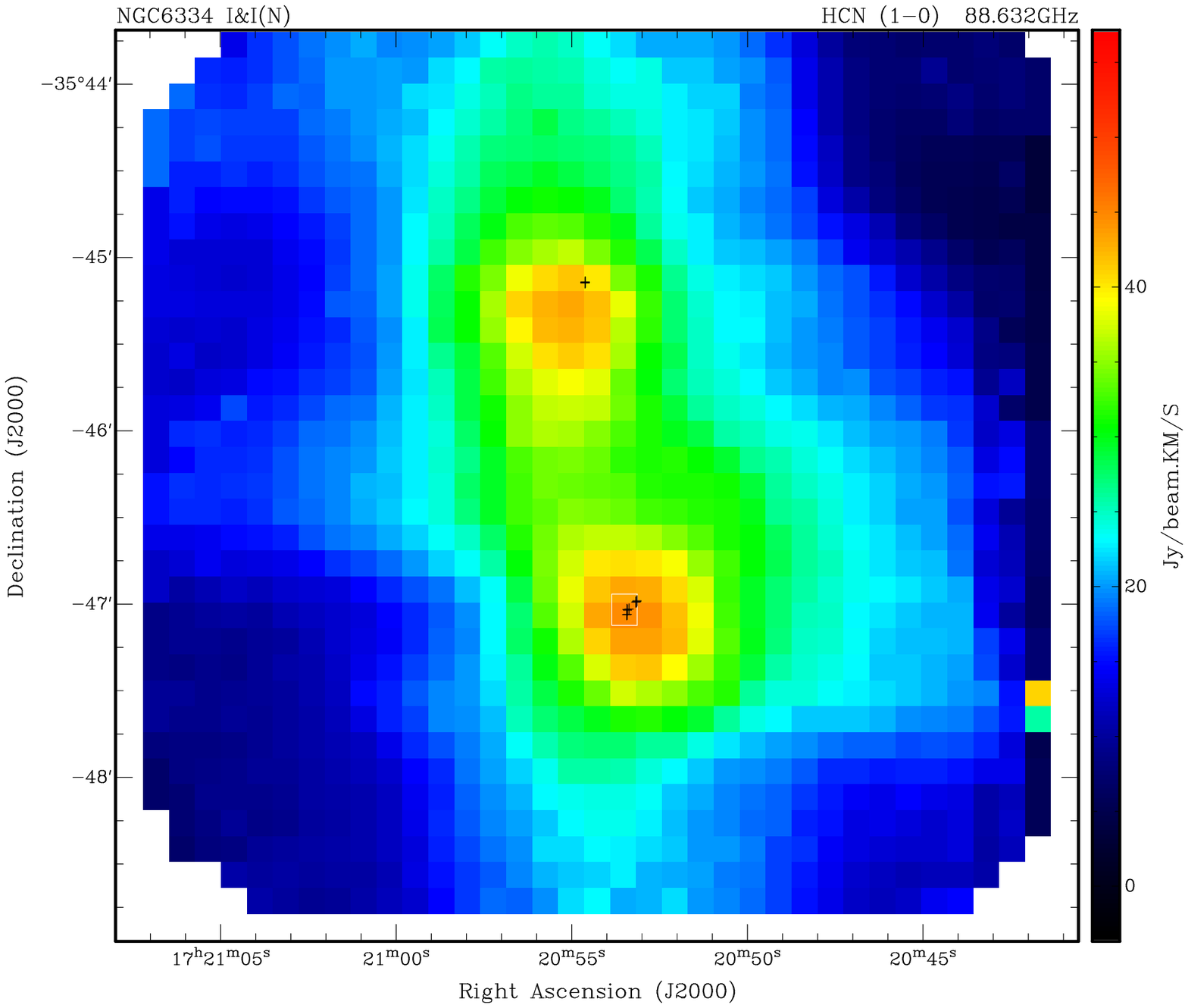}  
    \end{center}    
  \end{minipage}
  \hspace*{1.1cm}
  \begin{minipage}[t]{3cm}
\begin{center}
\hspace*{-.75cm}  
      \includegraphics[scale=0.225]{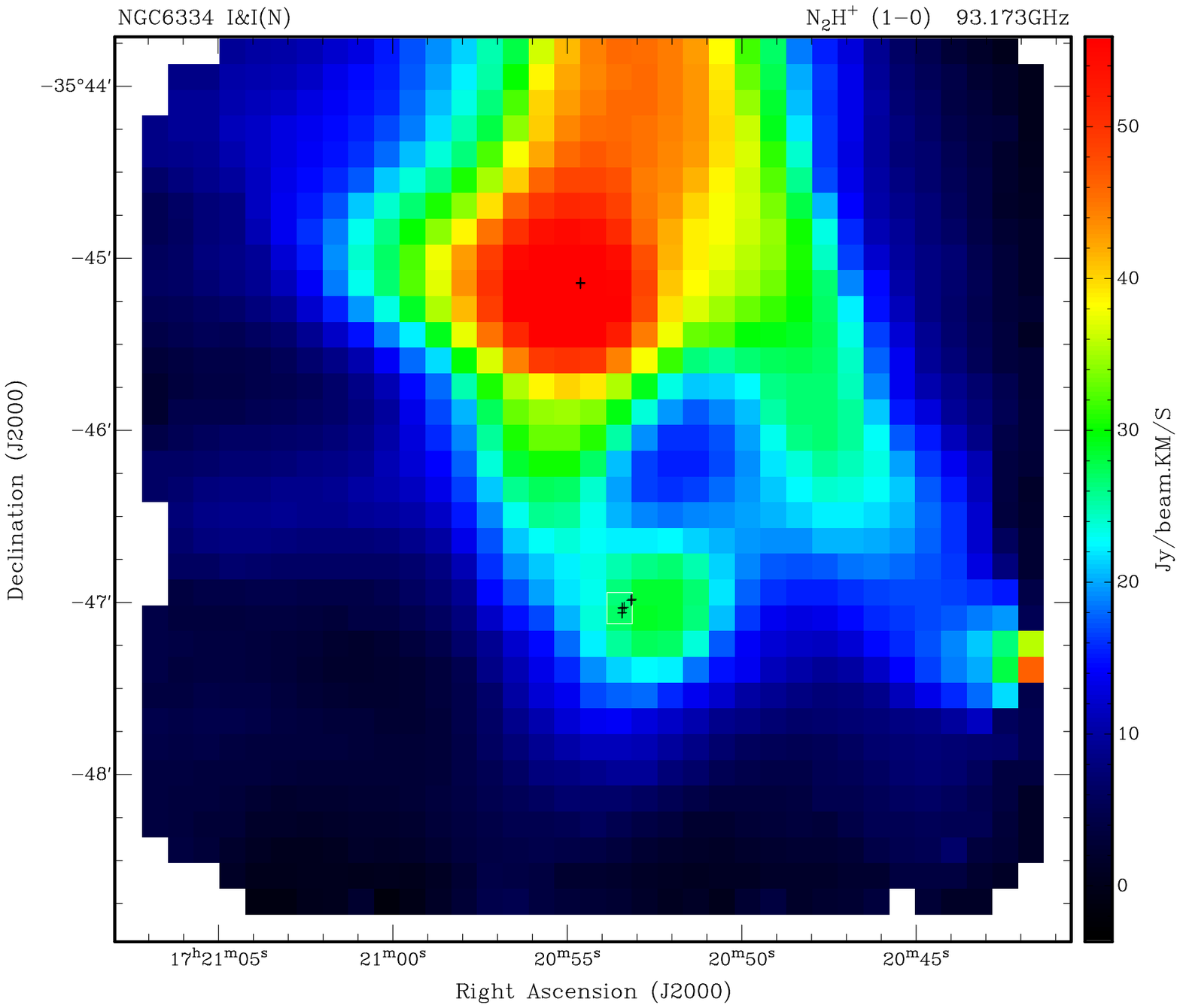}  
    \end{center}    
  \end{minipage}
  \begin{minipage}[t]{12cm}
\begin{center}
    \caption{Mopra maps of $^{13}$CO($1-0$) (left), HCN($1-0$) (center), and N$_2$H$^+$($1-0$) (right) towards the regions NGC6334 I (southern core seen in HCN) and I(N) (northern core) displaying the morphology of the region as seen in different molecular tracers.}
\end{center}
  \end{minipage}
 
\end{figure}



\end{document}